**Innovative System Design for Remote Air Traffic Control Simulation Training**

Liang, Man

RMIT STEM College, RMIT University, Melbourne 3001, Victoria, Australia

**Abstract**

Conventional Air Traffic Control (ATC) simulation training is a stand-alone system. Trainees are required to be in the simulation room to improve their practical skills. COVID pandemic challenges the current system and needs a new design to feed remote teaching and training. This research paper aims to introduce an innovative system design to support a sustainable ATC education on and beyond COVID.

The proposed remote ATC simulation training involves two main parts, namely i) a social-distancing on-campus simulation lab for students and ii) teaching by remote ATC instructors across States. We mirror the students' real-time practice scenario on a remote computer screen that a remote ATC instructor can observe. At the same time, the ATC instructor can communicate with the students to help them improve their practical skills. There are several benefits associated with the application of remote ATC simulation training. Firstly, the remote instructor is not required to be present on campus, which satisfies COVID social distancing and contactless requirements. Secondly, teaching ATC skills requires active engagement with ATC instructors who provide high quality and immediate feedback to students, resulting in a better learning experience. Thirdly, blended face-to-face and remote education innovation align with the future of education, as it is more flexible and dynamic and improves cost-efficiency. However, the remote ATC simulation training also poses challenges to universities and ATC training service providers. These challenges include the need for infrastructure design to feed the existing on-site training materials and software network design that considers cyber security risks due to online data sharing.

Remote ATC simulation training is an emerging technology in aviation education during COVID. Professional training institutions can learn from others, whereas the experiences of developing remote ATC simulation teaching/training with the Start-up company ByteProTeq will be beneficial for the rest of the world to understand the differences and similarities of current remote training, and to improve their performance in building a safe and efficient remote training environment. In this paper, we will present three improvements to our remote ATC training: 1) infrastructure upgrading of hardware and software from an existing stand-alone system to a remote network that considers costs, cyber security, system compatibility, et cetera; 2) quality of remote ATC simulation training, compared with traditional face-to-face training, including students' and instructors' feedback; 3) enhancement of the current remote training system beyond COVID-19 regarding reliability, cyber security, and capacity. This foundation paper will support understanding of the current stage of remote ATC simulation training development with the Start-up company ByteProTeq, during and beyond the COVID-19 pandemic, thereby providing an excellent example for the rest of the world.

## 1. Introduction

Since COVID-19 travel restrictions were announced in Australia on 1 Feb 2020 for travellers from mainland China, until the date being, there have been a massive number of international students forced to stay in their own countries. Significant parts of international students are from China and India, which accounted for 58% of higher education enrolments in 2019. University finance data, as reported by DESE shows that fees from overseas student enrolments accounted for 26% ($8.8 billion) of university revenue in 2018. Universities that highly rely on international students' prices have fallen heavily during COVID-19. According to the report published on 3 Feb 2021 by Universities Australia, Australian universities shed at least 17,300 jobs in 2020 and lost an estimated $1.8 billion in revenue compared to 2019. It is estimated to lose a further 5.5 per cent, or $2 billion, in 2021. On the other hand, onshore students still have difficulties physically attending lectures and tutorials due to periodic social distancing restrictions.

Australian universities must think of more solutions to manage the courses resources and offer more options for students to ensure the number of enrolments. As a result, education mode is quickly changing and adapting to the pandemic challenges. The rise of online/remote teaching and learning during the COVID-19 pandemic challenges all levels of education in different disciplines. In the aviation field, conventional Air Traffic Control (ATC) simulation training is being transformed into a remote training mode in higher education institutions and international ATC training providers.

Researchers found that the COVID-19 pandemic forces people to make large-scale behavioural changes, placing significant psychological burdens on individuals [1]. In the education field, there are a lot of unknowns in the remote teaching mode. For example, large class lectures are offered online for students during social distancing for both onshore and offshore students. Instead of facing hundreds of students in the lecture theatre, the lecturer delivers the course via zoom in front of one or several computer screens. The intense feeling of self-talking is a dramatic change for lecturers. On the other side, the students could see the lecturer virtually online. However, the virtual teaching environment currently offered by zoom also makes students be isolated from their colleagues physically, and few social processes could be engaged quickly.

As mentioned above, the proposed remote ATC simulation training in this paper involves two main parts, namely i) a social-distancing on-campus simulation lab for students and ii) teaching by remote ATC instructors across states. We mirror the students' real-time practice scenario on a remote computer screen that a remote ATC instructor can observe. At the same time, the ATC instructor can communicate with the students to help them improve their practical skills. There are several benefits associated with the application of remote ATC simulation training. Firstly, the remote instructor is not required to be present on campus, which satisfies COVID social distancing and contactless requirements. Secondly, teaching ATC skills requires active engagement with ATC instructors who provide high quality and immediate feedback to students, resulting in a better learning experience. Thirdly, blended face-to-face and remote education innovation align with the future of education, as it is more flexible and dynamic and improves cost-efficiency. However, the remote ATC simulation training also poses challenges to universities and ATC training service providers. These challenges include the need for infrastructure design to feed the existing on-site training materials and software network design that considers cyber security risks due to online data sharing.

ATC simulation is an emerging technology and aviation education. Professional training institutions can learn from others, whereas the experiences of developing remote ATC simulation teaching/training in this paper will be beneficial for the rest of the world to understand the differences and similarities of current remote training, and to improve their performance in building a safe and efficient remote training environment.

## 2. Innovative system design for remote ATC training system

### 2.1 Pre-COVID stand-alone ATC training system

Air traffic controllers manage the flow of aircraft into and out of the airport airspace, guide pilots during take-off and landing, and monitor aircraft as they travel through the skies. Their primary concern is safety, but they must also direct aircraft efficiently to minimize delays. According to the standards and recommended practices set by the International Civil Aviation Organization (ICAO), air traffic controllers must meet several specific requirements regarding their age, knowledge, experience, and medical fitness. For example, they need good spatial awareness and strong mathematical skills, excellent communication skills, the ability to work well under pressure and make quick, accurate decisions, and the capability to plan and adapt to changing situations [2].

The training of air traffic controllers can be divided into several defined phases that cover both basic and advanced training. Pre-COVID ATC training worldwide is mainly delivered in face-to-face mode, mostly consisting of a ratio of one instructor to one student. The training includes both theory and practical components. Practical training is essential for student air traffic controllers. The practical training is undertaken through an ATC simulator, i.e. a tool that provides a realistic imitation of the environment a controller works in. The ATC Simulator is a stand-alone system designed to provide a high-fidelity environment in which air traffic controllers can be trained from student controller through to rating and conversion and specialist applications such as emergency and unusual situations. Figure 1 illustrates the typical framework of a "stand-alone ATC simulator with two-position hardware configurations". LAN is a local area network that is used to connect all types of stations in this system. On the bottom, there are Pseudo Pilot stations and two Controller Student stations. On the top there are Supervisor station, Host station and Excise editing station.

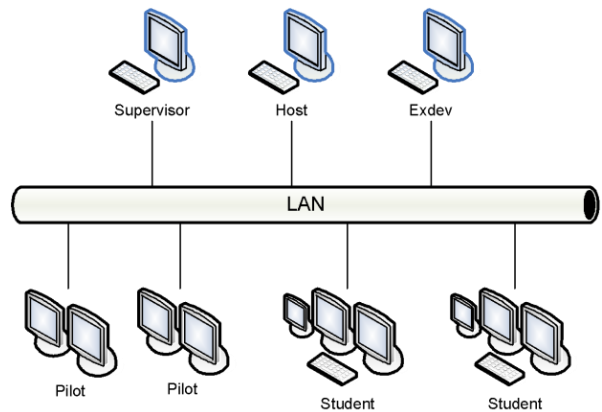

Figure 1 Stand-alone ATC simulator with two-position hardware configurations

At the University of South Australia (UniSA), a Pre-COVID, stand-alone ATC simulator system is used to help international and domestic students in Australia to explore an ATC career in the aviation industry, to strengthen their passion for aviation and to understand why and how an air traffic control system works. All assessment items for the course AERO 3019 Air Space Management are designed to support students to develop knowledge, understanding and practical skills about air traffic control. ATC Simulation scenarios are designed for students to apply their knowledge in practical situations. At UniSA, the ratio of students to instructors in the ATC simulation room is 6 to 1, which means that one instructor must observe two ATC stations and two pilot stations. Four students on two ATC stations are doing a simulation exercise: one student acts as a controller and the other as a coordinator. Two students are on pilot stations, and they are responsible for following ATC instructions and updating aircraft trajectories. As shown in Figure 3 and Figure 4, the ATC simulator in the UniSA ATC simulator room is a stand-alone system that only contains the student controller station, pseudo pilot station, and supervisor station [3].

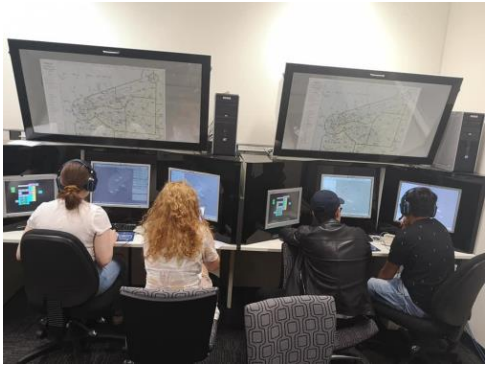
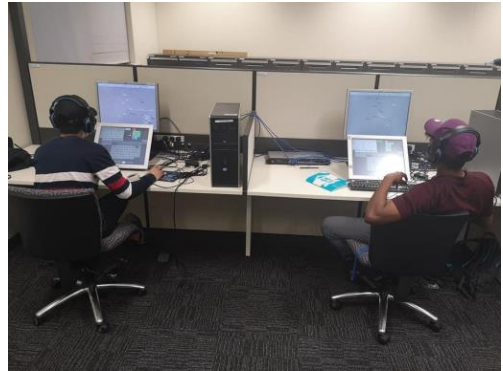

Figure 3 Student controller role station            Figure 4 Pseudo pilot station

In Pre-COVID, the maximum capacity of this course is 30 students per semester. There is a maximum of 6 students in each session at one time. We currently only have two stations that are operational and one instructor on site. As a result, five sessions must be scheduled for different groups of students. In each session, students need to rotate to perform the controller's role on controller position. Each student has 20-30 minutes of practice time. As the exercise lasts typically 1 hour, students are not always doing the same exercise with their colleagues, and the practice time for each student is limited. Usually, the first student doing the exercise feels much more challenged than the others, as they need more effort to build up situational awareness.

**2.2 Remote ATC training system development on COVID-19**

The nature of ATC skills training involves many practical exercises with qualified instructors. Thus, simulation courses play an essential role in training air traffic controllers. However, the availability of instructors or Pseudo pilots could be a challenge for some aviation academies. It isn't easy to offer students the best opportunity for one-on-one, face-to-face teaching. Previous researchers investigated current and future uses of simulation in the Federal Aviation Administration (FAA) Academy's ATC training program [4]. Their primary findings emphasize the importance of applying new technologies such as web-based training methods to reduce the dependence on on-site instructors during simulation training. Other researchers, such as Dönmez, Demirel and Özdemir, solved the Pseudo-pilot assignments problem via Mixed-integer programming (MIP), by considering the availability Pseudo-pilot in terms of period, personal mental or physiological workloads [5].

Remote ATC simulation is an emerging technology in aviation education. At the beginning of 2020, a teaching innovation grant was used to support the development of a small ATC tutoring project at UniSA. The principal aim of the teaching innovation project is to improve the students' learning experience via remote tutoring to offer students the one-on-one skills training experience. The 'Remote ATC tutoring tool' or the 'E-tutoring tool' aims to 1) lower the ratio of students to instructors by allowing more remote instructors to engage with students in the ATC practical exercises; 2) improve students' learning experience in remote learning mode. In this project, there are limited financial resources available. Therefore we need to consider the cost, timeline, cyber security, and system compatibility with the university-wide network. In the end, we mainly focus on two changes: 1) infrastructure upgrading of hardware and software from an existing stand-alone system to a remote system compatible with university network; 2) remote teaching course content design to assure a high quality of remote teaching. There are several challenges in the remote system design and implementation: Firstly, there is limited finance available for this project, redesigning the whole ATC simulation system is impossible. Therefore a plug-in implementation mode is considered to upgrade the software and hardware to support a virtual remote teaching environment in ATC simulation training. Some applications such as Microsoft Teams could be plugged in and support a remote teaching environment with multiple functions, however, it required access into a vast network. Secondly, the standard university computer operating system is based on Windows, while the ATC simulation system is based on Unix. How to make two operating systems work together to offer virtual remote ATC training under budget is essential. There are several solutions, either via hardware upgrading or software upgrading, the costs are different. In addition, universities only offer standard IT hardware for hardware upgrading. The remote ATC simulation system must be compatible with university IT standards and requirements. Thirdly, future development should be considered. Modules and cloneable solutions will be much convenient.

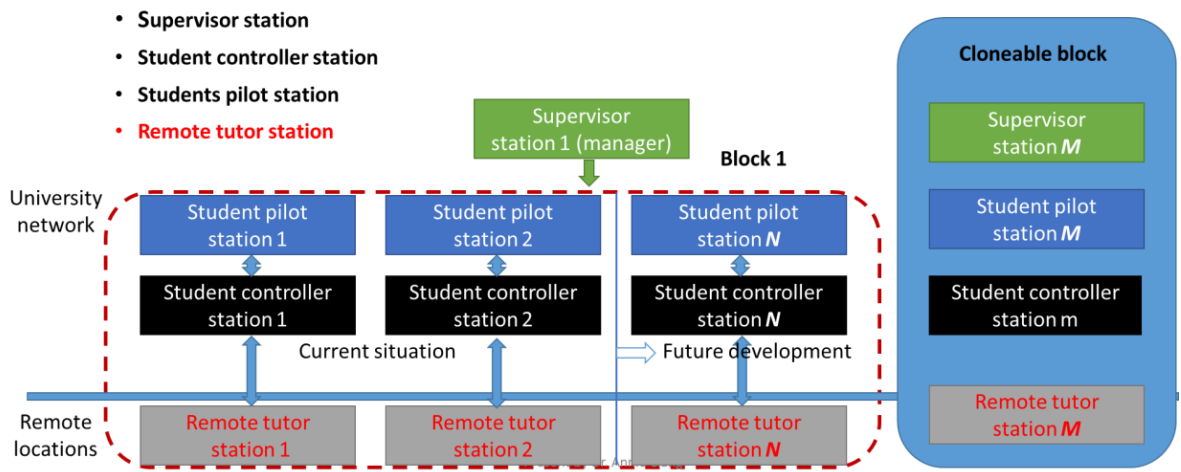

Figure 5 Remote ATC simulation training system framework

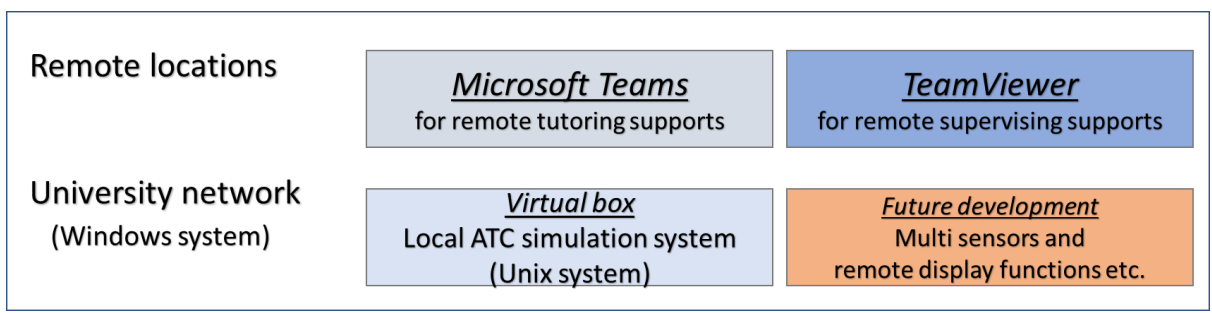

Figure 6 Plug-in teaching software design for Remote ATC simulation training system

In the end, the remote system framework and Plug-in software network design are shown in Figure 5 and Figure 6. In the framework, for each student controller, there is a remote tutor to support the learning. The remote tutor could be anywhere in the world if they have access to the internet. Student pilots and student controllers collaborate in the simulation room. These two stations are geographically separated from each other in isolation. One supervisor could manage up to 10 student controllers and ten student pilots simultaneously, named "a block". If there is an extension in the future, then the cloneable block could be produced.

In the university network, all stations are running under Windows operating systems which is a university standard. If there is a problem with the computer box, the university IT support team could replace it quickly. Then inside the window system, there are 1) virtual box to run the local ATC simulation system with Unix operating system, 2) Microsoft teams (license required) to provide remote tutoring supports including virtual face contact, video/voice recording, remote control function from remote tutors, questionnaires etc., 3) Team viewer (license required) for supervising supports, 4) other sensor systems such for future development. In the remote location, the remote tutor should have a computer with Microsoft teams. That is the only requirement. Based on the system design, all the hardware needs to upgrade to university standards. Then there are challenges about handling four screens on the student controller station, synchronising two sound inputs from Windows and Unix systems into one, and double network interface cards to support the Local Area Network (LAN) and Wide Area Network (WAN). The software updating collaborates with Adelaide based IT Start-up company ByteProTeq. Finally, the upgraded stations Human Machine Interfaces (HMIs) are shown in Figure 7. In Figure 7(a), an additional screen on the top left for video communication will connect the remote tutor with student controllers. Using the new system, the simulation practices can be conducted without the need for the student and instructor to be at the same location. The students' real-time practice scenario can be mirrored to a remote computer screen, which a remote ATC instructor can observe. At the same time, the ATC instructors/tutors can communicate with the students to help them improve their practical skills. In addition, if remote instructors/tutors would like to demonstrate some functions in real-time, they could remote control the student controller station and show the correct actions. The remote instructor/tutor identification is shown in the student controller's main screen with red and circle shapes, see Figure 7 (a) right. On the Pseudo pilot station, students are not necessary to connect with remote instructors/tutors. They mainly communicate with student controllers. If they need any help, they could join Microsoft teams and request support from remote instructors/tutors (see Figure 7 (b). Meanwhile, with the support form application TeamViewer, the supervisor station could be remote-controlled, which offers the option to set up the exercise from any location in the states.

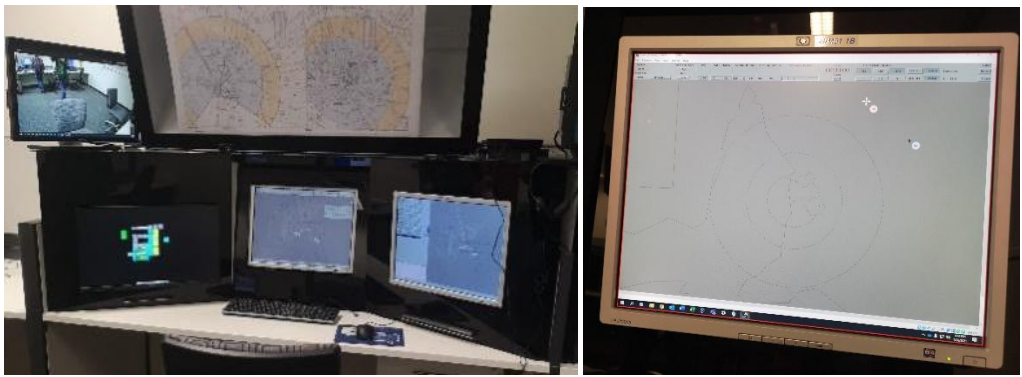

a) New student controller station HMI with four screens (Main screen with red colour and circle-shaped remote tutor identification, mirrored to remote tutor)

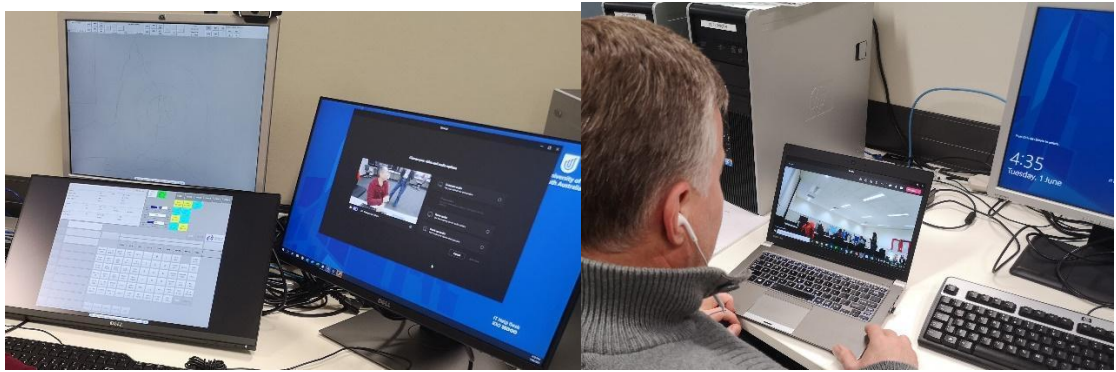

b) New student pilot station HMI (left side) and remote supervisor station HMI (right side)

Figure 7 New remote ATC simulation training system HMI

## 3. Preliminary experiments and discussions

### 3.1 Human-in-the-loop Experiments and survey results

Due to financial limitations, only three experiments were organized. The first experiment was organized on 1st June 2021. The second experiment was organized with 6 students involved on 17th June 2021 Thursday 9:30-11:30 am, and the last experiment finished on 24 June 2021 14:00-17:00 with the same students involved. All students in these experiments are familiar with old stand-alone systems and have some experience of initial ATC training. 4 of them have Private Pilot Licenses (PPL), 2 of them are aviation management students without PPL. 5 of them are male international students and only 1 female local student. They shifted their roles to be student controllers, student pilots, and remote tutors. After all experiments, surveys were distributed to all students to get the feedback about the quality of remote teaching and possible improvements of the system. The 10 questions are:

1) Are you comfortable with a remote tutoring environment?
2) How satisfied are you with the ability to integrate Remote tutor with on-campus ATC simulation training?
3) When you think about this remote ATC tutoring tool, do you think of it as something you need or don't need to enhance learning experience?
4) How satisfied are you with the look and accessibility of this Remote ATC simulation training tool?
5) Are you happy to provide remote tutoring service for new students in the airspace management course if you are qualified?
6) How many hours would you like to provide a continuous remote tutoring service per day if you are qualified?
7) If you have the right to choose your tutor, what is the most critical skill you are looking for?
8) Overall, are you satisfied or dissatisfied with the overall quality of the remote ATC simulation system?
9) One thing you liked most about the remote ATC simulation tutoring tool is?
10) One thing you thought could have been better or was missing is?

The results are listed in Figure 8 and Table 1. On average, all students are satisfied with the overall quality of the remote ATC simulation system. All students are comfortable with a remote tutoring environment for ATC simulation training and are very satisfied with integrating remote tutoring with on-campus ATC simulation training. It is a blended teaching/training mode for students. On one side, they could physically attend the simulation room and keep social distance from their colleagues. On the other side, the one-on-one tutoring support is from a remote tutor. The majority of Students well understand the need for remote tutoring systems on and beyond COVID-19. We are encouraged to keep all the education activities happening at universities as much as possible, so the accessibility of remote ATC simulation training systems is critical.

Remote ATC instructors or tutors are essential resources for the new system. By guiding a 1:2 ratio, students can better learn in an environment where their instructors/tutors can answer questions straight away. Qualification of remote ATC instructors/tutors is a crucial component in high-quality teaching. With different levels of trainees, there should be different levels of ATC instructors or tutors. More than ten years' experience ATC instructors are limited resources in South Australia. Still, if we would like to provide the initial ATC training for ab-initio students, then a pool of qualified ATC tutors could be trained by universities. Because in the experiments, most of the students are 3rd year international students and ready to graduate from Australian university. According to their feedback, they are very likely to provide the remote tutoring services for new ab-initio students and work for more than 2 hours per day to gain their working experiences in the aviation field. For open question 9 and 10 feedback, see Table 1; it is found that all the students are happy with instantaneous feedback from remote tutors. It is convenient to learn hands-on knowledge directly via a hidden mode. However, it is also found that operational manuals, communication devices, and remote tutor qualifications are significant concerns for the quality of remote ATC simulation training.

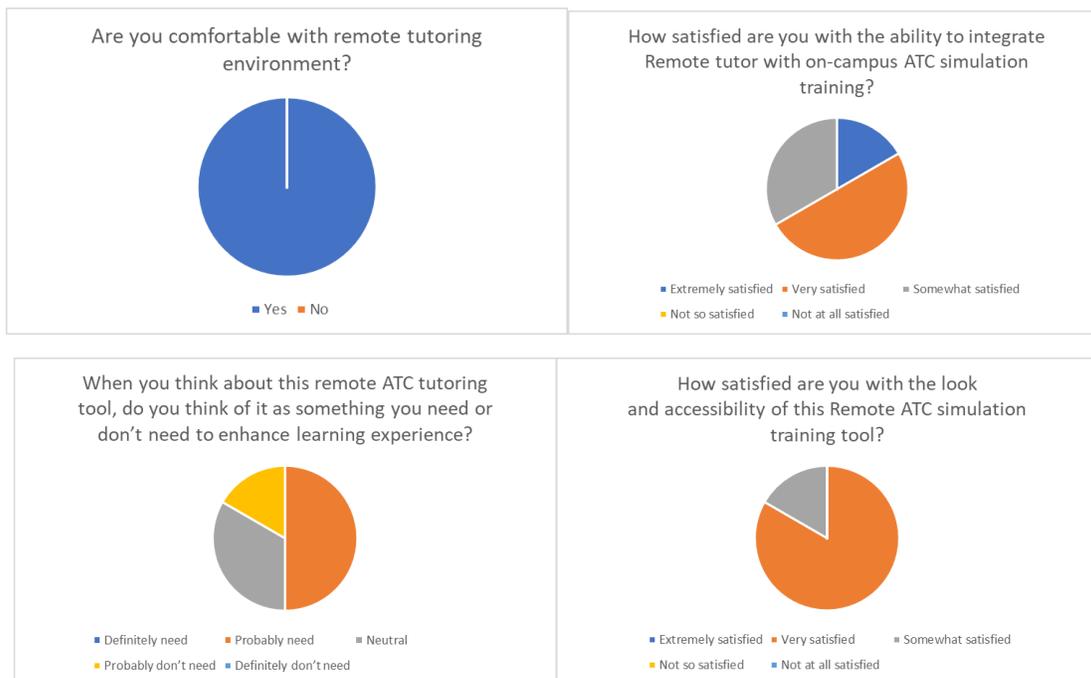

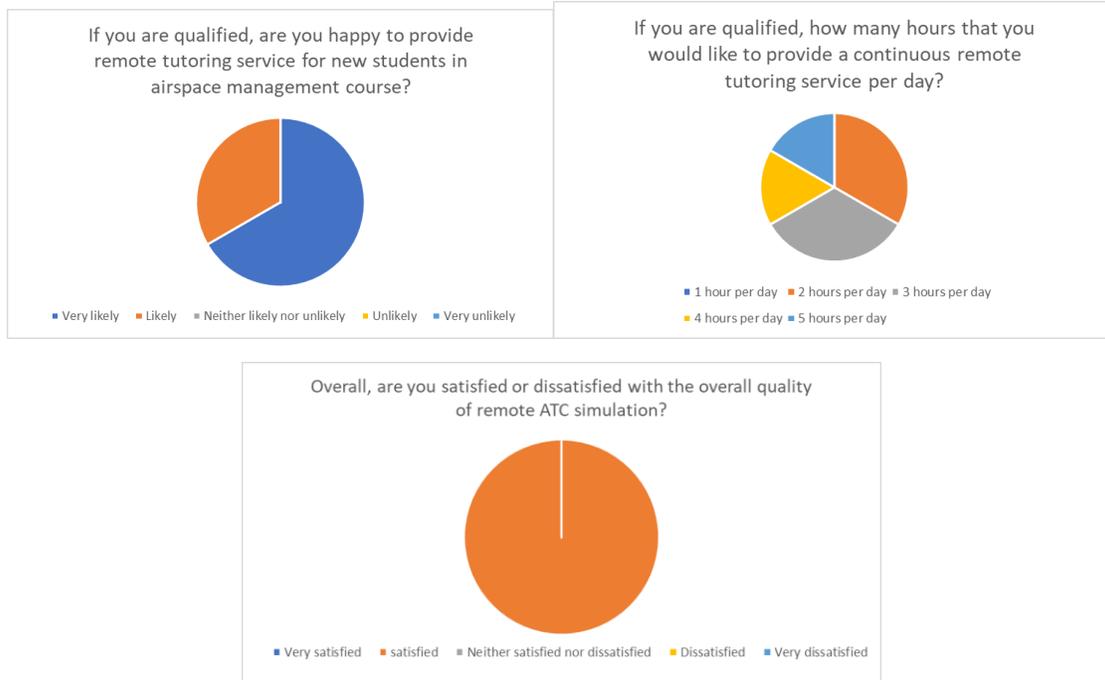

Figure 8 Survey results

Table 1 Open questions survey results

| Feedback | Question 9<br>**One thing you liked most about the remote ATC simulation tutoring tool is** | Question 10<br>**One thing you though could have been better or was missing is** |
|---|---|---|
| 1 | The ability to share screens with tutors will be invaluable when teaching. It will provide opportunities for instantaneous feedback and correction if necessary. | A detailed list of instructions on how to successfully establish communication. This should accompany a manual detailing operation of Microsoft Teams software. |
| 2 | The tutor can attend and supervise the class through Microsoft teams. | A way to integrate system sounds from both Unix system and Windows with a good quality. |
| 3 | Tutor can directly and remotely control the simulator through Microsoft Teams, which can offer students help whenever needed | The microphone's quality could be better. |
| 4 | The tutor doesn't need to show up, and since each tutor is going to instruct only one group, students will have more time to practice and can be better taken care of. | To make the remote tutor method work well, the preconditions are the "tutors are trained properly" and the "students really understand those basic control buttons". |
| 5 | Scalability of the concept. Remote tutoring provides tutors with the convenience of transferring knowledge to students from different locations. Also, by providing guidance in a 1:2 ratio, students can better learn in an environment where questions can be answered straight away by their tutors. | Troubleshooting procedures can be introduced to tutors by including in the tutor's manual. |
| 6 | Can observe the students operating straight from the monitor, correct them from mistakes by remote control. | Experiments with new students involved are recommended for further system test. |

### 3.2 Discussions

Emergency remote teaching soared during COVID-19. Higher education institutions are forced to make immediate changes from traditional face-to-face to more digital and remote mode. Conventional classroom lectures are replaced by virtual rooms using videoconferencing systems. At the same time, ATC simulation is more like a practical tutorial for students. Any video conferencing application could not simply replace it. A scenario-based ATC simulation training must be delivered via a blended teaching mode which consists of one part of teaching activities provided online and one part of learning activities happening on site. The learner's interaction with remote tutors via the Internet to learn, find out information, or do air traffic control strategies.

Online class is more challenging to handle than a traditional face-to-face class. A careful course design process should be organized in advance at least six months. Especially, user's manuals for student pilot stations, student controller stations, supervisor stations, and remote instructor/tutor stations should be clearly drafted, including troubleshooting procedures due to the high integration of technology. Remote tutors should get trained and qualified before providing the services.

The number of students demanded for this course was doubled in the last year from 30 to 60, and it is predicted to 100 in the coming teaching semester. Surveys also were widely distributed to students in Australia and offshore, the feedbacks was very positive. Therefore, enhancing the current remote training system beyond COVID-19 regarding reliability, cyber security, and capacity is also very important. Firstly, failure of the system could be malfunctions of hardware or software. Due to financial limitation, all used hardware is only functional, but not with high-performance requirements. A professional IT expert programs the software integration. The reliability of software systems is very high; only the hardware reliability needs to be validated in the future. Secondly, because of the remote ATC simulation system with video and voice recording functions, it provides online real-time teaching support. Thus the cyber security issue must be considered. Because the main computer boxes used in current remote ATC simulation systems are university Windows standards, the LAN of remote ATC simulation systems is under the umbrella of university WAN. Therefore, the cyber security level currently is handled on WAN level. Thirdly, the remote ATC simulation system is developed with a cloneable plug-in modular concept. It is very convenient to increase the capacity of the student controller stations to adapt to the demand, see Figure5.

### 4. Conclusion remarks

ATC simulation is an emerging technology and aviation education. Professional training institutions can learn from others, whereas the experiences of developing remote ATC simulation teaching/training with the Start-up company ByteProTeq will be beneficial for the rest of the world to best understand the differences and similarities of current remote training, and to improve their performance in building a safe and efficient remote training environment. In this paper, we have presented three improvements to our remote ATC training: 1) infrastructure upgrading of hardware and software from an existing stand-alone system to a remote network that considers costs, cyber security, system compatibility, et cetera; 2) quality of remote ATC simulation training, compared with traditional face-to-face training, including students' and instructors' feedback; 3) enhancement of the current remote training system beyond COVID-19 regarding reliability, cyber security, and capacity. The purpose of this foundation paper is to provide a better understanding of the current state of remote ATC simulation training development with the Start-up company ByteProTeq, during and beyond the COVID-19 pandemic, by giving an excellent example for the rest of the world and its leading role in developing advanced ATC training systems for application at a global level.